\documentclass[sigconf]{acmart}

\usepackage{color}
\usepackage{xspace}
\usepackage{listings}
\usepackage{subcaption}
\hypersetup{
  colorlinks=true,
  linkcolor=blue,
  citecolor=blue,
  urlcolor=blue
}
\usepackage{newfloat}

\definecolor{MyLightGray}{RGB}{230, 230,230}
\definecolor{MyDarkBlue}{RGB}{10, 10, 185}
\definecolor{MyCyan}{RGB}{20, 145, 145}
\definecolor{MyDarkGreen}{RGB}{0, 175, 60}
\definecolor{MyDarkRed}{RGB}{175, 0, 0}
\definecolor{MyBrown}{RGB}{165,42,42}
\newcommand{\CodeSymbolGreen}[1]{\textcolor{MyDarkGreen}{#1}}

\newcommand{\CodeSymbolCyan}[1]{\textcolor{MyCyan}{#1}}
\newcommand{\CodeSymbolRed}[1]{\textcolor{MyDarkRed}{#1}}
\newcommand{\CodeSymbolBrown}[1]{\textcolor{MyBrown}{#1}}

\newcommand{\resourceName}{Q\texorpdfstring{$^2$Forge\xspace}\xspace}
\newcommand{\genkgbot}{Gen\texorpdfstring{$^2$KGBot\xspace}\xspace}
\newcommand{\qset}{Q\texorpdfstring{$^2$set\xspace}\xspace}
\newcommand{\qsets}{Q\texorpdfstring{$^2$sets\xspace}\xspace}

\AtBeginDocument{%
  }

\lstdefinestyle{sparqlStyle}
{   basicstyle=\scriptsize\ttfamily,breaklines=true,
    numbersep=3pt, xleftmargin=0.3cm, xrightmargin=0cm,
    columns=fullflexible,
    keywordstyle=\color{red},
    keywords=[1]{@prefix,PREFIX,@base,BASE,SELECT,DISTINCT,ORDER,BY, VALUES, FILTER, WHERE,UNION, GROUP, BIND, INSERT, ASK, SERVICE},
    morekeywords=[2]{{[},{]}},
    morekeywords=[2]{{\{},{\}}},
    breaklines=true,
    comment=[l]{\#},
    morecomment=[s][\color{blue}]{<}{>},
    tabsize=4,
    alsoletter={-?}, 
    morestring=[b][\color{black}]\",
    showstringspaces=false,
     literate= {;}{{\CodeSymbolRed{;}}}1
     {.}{{\CodeSymbolRed{.}}}1
     {"}{{\CodeSymbolRed{"}}}1
     {\{}{{\CodeSymbolGreen{\{}}}1
     {\}}{{\CodeSymbolGreen{\}}}}1
     {]}{{\CodeSymbolCyan{]}}}1
     {[}{{\CodeSymbolCyan{[}}}1
     {(}{{\CodeSymbolBrown{(}}}1
     {)}{{\CodeSymbolBrown{)}}}1,  
    moredelim=[s][\color{MyDarkBlue}]{:}{\ },
    moredelim=[s][\color{MyDarkRed}]{@}{\ },
}

\lstdefinestyle{tupleStyle}
{   basicstyle=\scriptsize\ttfamily,breaklines=true,
    numbersep=3pt, xleftmargin=0.3cm, xrightmargin=0cm,
    columns=fullflexible,
    breaklines=true,
    comment=[l]{\#},
    tabsize=4,
    alsoletter={-?}, 
    showstringspaces=false,
}

\DeclareFloatingEnvironment[
  fileext=lol,
  name=Multiple Listing
]{listing}
\DeclareCaptionSubType{listing}

\begin{document}

\title{\resourceName: Minting Competency Questions and SPARQL Queries for Question-Answering Over Knowledge Graphs}

\author{Yousouf Taghzouti}
\email{yousouf.taghzouti@univ-cotedazur.fr}
\orcid{0000-0003-4509-9537}
\affiliation{
  \institution{Univ. Côte d'Azur, Inria, ICN, I3S}
  \city{Nice}
  \country{France}
}

\author{Franck Michel}
\email{franck.michel@inria.fr}
\orcid{0000-0001-9064-0463}
\affiliation{
  \institution{Univ. Côte d'Azur, CNRS, Inria, I3S}
  \city{Nice}
  \country{France}}

\author{Tao Jiang}
\email{tao.jiang@cnrs.fr}
\orcid{0000-0002-5293-3916}
\affiliation{
  \institution{Univ. Côte d’Azur, CNRS, ICN}
  \city{Nice}
  \country{France}
}
  
\author{Louis-Félix	Nothias}
\email{louis-felix.nothias@cnrs.fr}
\orcid{0000-0001-6711-6719}
\affiliation{
  \institution{Univ. Côte d’Azur, CNRS, ICN}
  \city{Nice}
  \country{France}
  }
  
\author{Fabien Gandon}
\email{fabien.gandon@inria.fr}
\orcid{0000-0003-0543-1232}
\affiliation{%
  \institution{Inria, Univ. Côte d'Azur, CNRS, I3S}
  \city{Nice}
  \country{France}}

\renewcommand{\shortauthors}{Y. Taghzouti et al.}

\begin{abstract}
The SPARQL query language is the standard method to access knowledge graphs~(KGs).  However, formulating SPARQL queries is a significant challenge for non-expert users, and remains time-consuming for the experienced ones. Best practices recommend to document KGs with competency questions and example queries to contextualise the knowledge they contain and illustrate their potential applications. In practice, however, this is either not the case or the examples are provided in limited numbers. Large Language Models (LLMs) are being used in conversational agents and are proving to be an attractive solution with a wide range of applications, from simple question-answering about common knowledge to generating code in a targeted programming language. However, training and testing these models to produce high quality SPARQL queries from natural language questions requires substantial datasets of question-query pairs. In this paper, we present \resourceName that addresses the challenge of generating new competency questions for a KG and corresponding SPARQL queries. It iteratively validates those queries with human feedback and LLM as a judge. \resourceName is open source, generic, extensible and modular, meaning that the different modules of the application (CQ generation, query generation and query refinement) can be used separately, as an integrated pipeline, or replaced by alternative services. The result is a complete pipeline from competency question formulation to query evaluation, supporting the creation of reference question-query sets for any target KG.
\end{abstract}







\maketitle

\section{Introduction}
\label{sec:1}

Semantic technologies, and in particular knowledge graphs (KGs), have been utilised in a variety of applications over time, including search engines, data integration, enterprise settings and machine learning. Numerous methods were proposed to assist their life-cycle and exploitation~\cite{hogan2021knowledge} leading to their adoption and the rapid growth of the Linked Open Data (LOD) cloud.\footnote{Statistics on the Linked Open Data cloud: \url{https://lod-cloud.net/\#about}}
However, the exploitation of these KGs has been hindered by the steep learning curve associated with the stack of standards, in particular query languages such as SPARQL~\cite{warren_sparql:2020}.

Over the last few years, common information retrieval methods have been profoundly renewed by the emergence of pre-trained Large Language Models (LLMs). The abilities of LLMs to understand and generate natural language (NL) and code alike have opened new research and development fields notably in the domain of data access and interaction.
In particular, these abilities endow LLMs with the ability to translate a question expressed in NL into its counterpart in a structured query language, SPARQL in the case of RDF KGs.
This allows domain experts to ``speak to structured data'' thus facilitating data access.
To design and evaluate such text-to-SPARQL translation systems effectively, we need reference datasets providing curated question-query pairs that are either tailored to a specific KG or at least relevant for the domain it concerns.

Some question-query datasets (that we hereafter refer to as \textbf{\qsets}) have been produced in the context of benchmarks and challenges such as QALD~\cite{QALD10}, DBNQA~\cite{DBNQA}, and LC-QuAD~\cite{LcQuAD}, but they are mostly based on subsets of DBpedia and/or Wikidata. 
When it comes to other domain-specific, possibly private KGs, or highly specialized KGs like in life sciences, creating a \qset involves skills that are rarely mastered by one and the same person.
More likely, this requires the collaboration of domain experts who can think of possibly complex competency questions (CQ) that scientists may want to ask, and Semantic Web experts who shall leverage the used ontologies and KG schema to come up with counterpart SPARQL queries.

Besides, a good practice in terms of documentation and metadata is to publish KGs with examples of queries they support. Yet, in practice this is rarely the case.
Similarly, whereas CQs have been identified as a valuable documentation and starting point for understanding the capabilities of a KG, many KGs are accompanied with very few CQs, if any at all.

To support the creation of \qsets aimed at training, testing, benchmarking, and documenting our systems and knowledge graphs, we identified the need to provide tools that help researchers--as well as scientific and technical information professionals--to understand existing KGs and generate or refine corresponding \qsets, whether they are Semantic Web newcomers or experienced practitioners.
Various methods and tools exist to help to create CQs and equivalent queries~\cite{alharbi_cq:2025,ciroku_cq:2024,pan_rag:2025,zouaq_schema:2020,rebboud_llm:2025,cohen_sparql:2013,emonet_llm:2024}. However, to the best of our knowledge, these tools are either domain-specific, extensively manual, or address only specific steps, but do not provide an end-to-end, integrated pipeline.

In this paper, we address the needs described above by presenting the methods, tools and services implemented in \textbf{\resourceName, a web application guiding the user through the steps of a generic, extensible, end-to-end pipeline to generate a reference \qset, i.e. a dataset of (NL question, SPARQL query) pairs tailored to a specific KG.}
Through an interactive and iterative process, the user interface assists the user in three main areas:
(1) producing CQs based on information about a KG and the domain it pertains to;
(2) proposing \textbf{SPARQL query counterparts of the CQs}, given the KG and its schemata;
(3) \textbf{testing} the proposed SPARQL queries, \textbf{judging} the relevance of question-query pairs, and \textbf{recommending refinements}.
Rather than constraining users to a fixed end-to-end pipeline, \resourceName emphasizes flexibility by allowing users to use one task independently of the others.

\resourceName relies on an extensive, user-controlled configuration where, in particular, multiple language models can be selectively used at different steps of the pipeline.
Through a documented Web API, \resourceName leverages a set of pre-defined services, e.g. to explore the KG or invoke a language model for a certain task, implemented using robust, community-proven libraries and frameworks such as LangChain.\footnote{LangChain homepage: \url{https://www.langchain.com/}}
Yet, a community may easily extend \resourceName with new services and steps, or re-implement some of the provided services for instance to use their own text-to-SPARQL tool instead of the one provided.

This paper is structured as follows: 
Section~\ref{sec:relatedworks} provides an overview and comparison with relevant existing works.
Section~\ref{sec:pipeline_descr} describes our methodology, the pipeline architecture and the various components. 
Then, the source code and its sustainability plan are described in Section~\ref{sec:code_doc}.
Section~\ref{sec:use_case} discusses real practical use cases where the resource could be used, while its potential impact and reusability are discussed in Section~\ref{sec:impact_reuse}.
Finally, the limitations and perspectives of the resource are outlined in Section~\ref{sec:conclusion}.


\section{Related Work}
\label{sec:relatedworks}
\paragraph{Linked Data Query Assistants.} Approaches for assisting users in querying KGs can be broadly separated in two main non-disjoint categories: the ones relying on dedicated Graphical User Interfaces (GUI) (e.g.~\cite{ferre:hal-01485093,10.1007/978-3-031-43458-7_2,10.1145/3511808.3557093}) and the ones relying on Natural Language Interfaces (NLI) (e.g.~\cite{lehmann:hal-04269089,tysinger:hal-04381448}). 
GUIs can provide high expressivity but remain difficult to use by non-technical experts, unless they trade off part of the expressivity in favor of reusing a popular interaction paradigm (e.g. faceted search). 
NLIs range from keyword-based retrieval to controlled natural language and full natural language dialogical interactions. Language models, large (LLM) and small (SLM), have significantly improved the methods for natural language processing in general, and in particular for question-answering over linked data. Language models are used in several ways: translate a question to a structured query (directly or indirectly \cite{lehmann:hal-04269089}) or by directly answering the question when, for instance, the knowledge source was included in the training corpus of the language model. 
These two trends can be combined with augmentation techniques such as Retrieval Augmented Generation (RAG)~\cite{lewis_rag:2020} that performs information retrieval tasks of different natures (document, database, KG) to enrich the context used to invoke the language model and improve the quality of the answers. While some GUIs help lower the barrier for non-expert users, NLI approaches, and particularly those using LLMs, are even more user-friendly and extensible but have shown mixed results in generating accurate SPARQL queries from natural language, and they require reference \qsets to be trained, augmented and evaluated. This is precisely the purpose of \resourceName.

\paragraph{Linked Data Question-Query sets.} 

Multiple tools and services are currently being released to address the question of text-to-SPARQL translation.
BigCQ~\cite{bigCQ:2022,bigCQ:2021} aims to create CQs and SPARQL equivalents based on the axioms of a specific OWL ontology. It is meant to help ontology engineering and evaluation, but cannot apply to common KGs that typically rely simultaneously on multiple ontologies and vocabularies.
Amazon Bedrock\footnote{\url{https://tinyurl.com/z4wrxvb3}} is a commercial, text-to-SPARQL translation service proposed by Amazon, that leverages LLMs and requires a collection of few-shot question-query pairs. It has applications particularly in bioinformatics.
Unfortunately it is not released under an open source license, thus making comparison difficult.
AllegroGraph's Natural Language Query (NLQ)\footnote{\url{https://franz.com/agraph/support/documentation/nl-query.html}} vector Database stores pairs of NL questions and corresponding SPARQL queries. This repository helps to train and refine models for accurate query generation, and its integration with SHACL shapes ensures the structural validity of the generated queries. However, this solution suffers from a lack of explainability. Users receive SPARQL query counterparts without understanding why a particular result was returned or the full process used to infer that outcome.
In the opposite direction, AutoQGS~\cite{10.1145/3511808.3557246} is a framework that generates NL questions from SPARQL queries, facilitating the creation of question-query datasets without extensive manual annotation. However, this solution requires existing SPARQL queries to generate training data, which limits its applicability.

Challenges such as QALD~\cite{QALD10}, DBNQA~\cite{DBNQA}, and LC-QuAD~\cite{LcQuAD} provide \qsets to train models that generate queries from NL questions, but they focus primarily on DBpedia and Wikidatan, although some editions of QALD, e.g. QALD~4\footnote{\url{https://github.com/ag-sc/QALD/tree/master/4/data}}, have included biomedical \qsets. Similarly, the LC-QuAD~2.0 dataset\footnote{\url{https://github.com/AskNowQA/LC-QuAD2.0/}} contains a \qset of more than 20,000 pairs across DBpedia and Wikidata, including subdomains such as geography and science.
While these resources serve general purpose question answering systems, they do not comprehensively capture domain specific or highly specialised knowledge. Some domain specific datasets exist, such as SIB bioinformatics SPARQL queries~\cite{bolleman_sib:2024}, a collection of hand-crafted \qsets for various SIB-related KGs. By contrast, \resourceName aims to fill this gap by providing an open-source solution to generate \qsets for any domain and KG, including for private KGs.


\section{From a Knowledge Graph to \resourceName Pipeline}
\label{sec:pipeline_descr}

\resourceName helps users to carry out three main tasks:
generate competency questions in NL for a target KG, generate SPARQL query translations of the competency questions, and test and refine the SPARQL queries.
To do so, \resourceName orchestrates the use of various services to manage multiple per-KG configurations, extract and pre-process the schema of a KG, invoke various language models depending on the task to achieve at each step of the pipeline, etc.
The services are invoked through a documented Web API implemented by a back-end server. We provide a prototype implementation of the back-end server called \genkgbot.
A community may reuse \genkgbot as-is, or they may customize or extend its services to meet their specific needs.
The links to the source repositories are given in Section~\ref{sec:code_doc}.

\begin{figure}
\centering
\includegraphics[width=\linewidth]{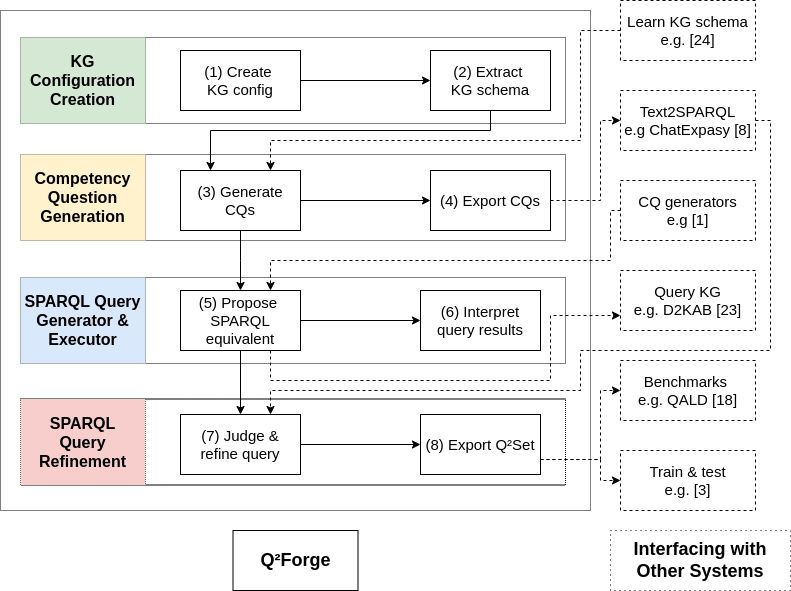}
\caption{\resourceName pipeline: resources and services.}
\label{fig:pipeline}
\end{figure}

The rest of this section further describes the steps of the \resourceName pipeline, that are depicted in Figure~\ref{fig:pipeline}:
(1)~create the configuration for a KG and 
(2)~extract and pre-process its schema;
(3)~generate CQs and 
(4)~optionally export them for reuse with another application or for documenting purpose; 
(5)~translate a CQ into SPARQL; 
(6) execute the query and propose an interpretation of the results; 
(7)~judge the relevance of the question-query pair and allow the user to iteratively refine the query; 
(8)~export the \qset for reuse with other systems.
Note that \resourceName remains very flexible: a user may follow the whole pipeline, but may also run each task independently by simply importing/pasting input data and exporting/copying the outputs.

\subsection{KG Configuration and Pre-processing}
\label{subsecsec:kg_config}

\subsubsection{Create a KG configuration.}
The pipeline starts with creating a KG configuration (depicted in Figure~\ref{fig:df_kg}) where the user provides minimal information about the target KG: a name, a short name used later as an identifier, a textual description, a SPARQL endpoint URL, and the namespaces and prefixes to be used in the SPARQL queries and Turtle descriptions.
Optionally, the user may fill in the URL of a SPARQL endpoint hosting the ontologies in case they are not on the same endpoint as the KG itself.

Once created, the configuration is stored on the back-end server.
Additional parameters that can be edited manually to configure the available language models (seq-to-seq and embedding), where they are hosted (e.g. local vs. cloud resources, vector database etc.), and how they are assigned to each step of the pipeline. 
For instance, one may choose to use a large model with reasoning capabilities for generating a SPARQL query, but use a smaller model to interpret SPARQL results.
Other parameters configure the strategy adopted to serialize ontology classes in the prompts submitted to seq-to-seq models, such as the number of ontology classes to describe and the linearization format used to describe them. Multiple formats are supported (currently Turtle, tuples or a NL format, see examples in Listing~\ref{list:embedding-cf}), since different language models may behave differently depending on the selected format.

\subsubsection{KG pre-processing.}
There is usually a gap between how an ontology defines classes and how instances of these classes are concretely represented in the KG. Typically, instances may use properties and resources from additional vocabularies that are not explicitly mentioned in the ontology.
Therefore, some downstream tasks like translating text to SPARQL and judging the relevance of a question-query pair require a description of the ontology classes as well as a description of how instances of these classes are concretely represented.

\begin{figure}[!t]
\centering
\includegraphics[width=\linewidth]{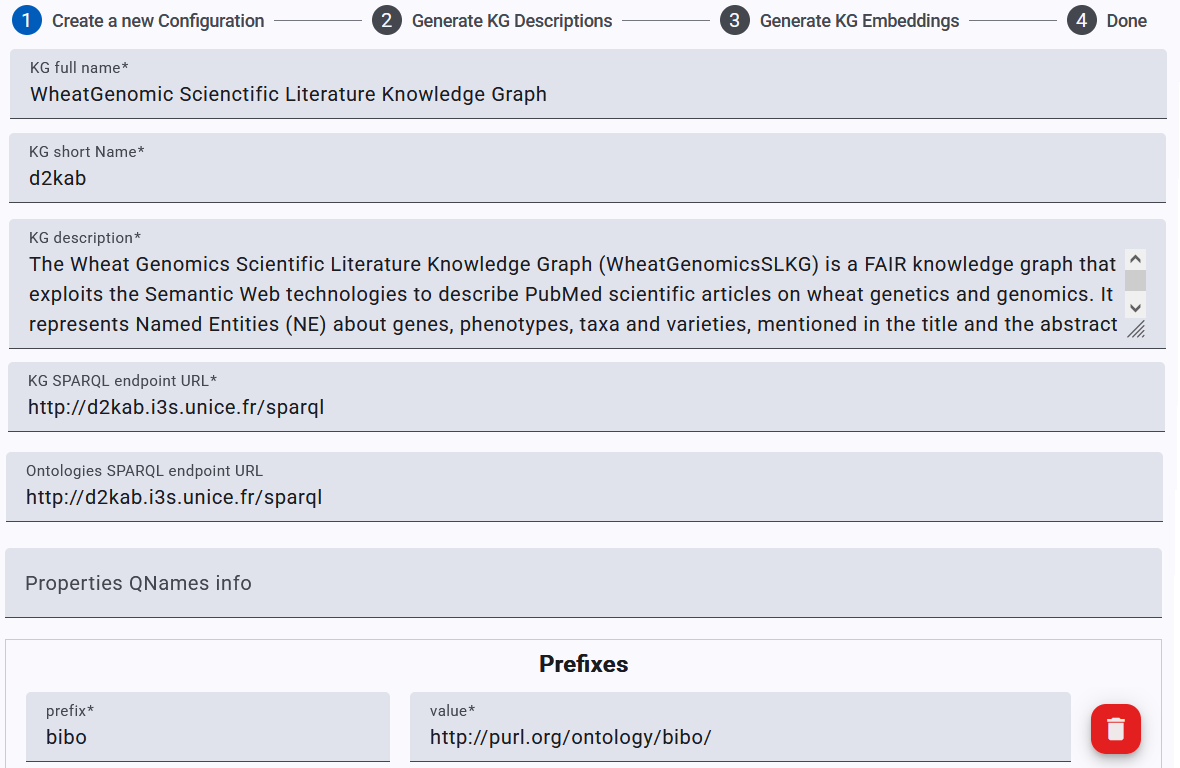}
\caption{The KG configuration and pre-processing interface.}
\label{fig:df_kg}
\end{figure}

To do so, we first extract from the KG various types of information that will be helpful to carry out the downstream tasks.
In our implementation, \genkgbot, this step creates a textual description of the classes from the labels and descriptions available in the ontologies, and computes text embeddings thereof.
In Figure~\ref{fig:df_kg}, these functions are available from the tabs 2 and 3.
Second, \genkgbot samples class instances and analyzes the properties and value types they use (examples are provided in Listing~\ref{list:embedding-cf}).
Lastly, the user may provide existing examples of NL question and associated SPARQL query. The pre-processing includes computing embeddings of these question-query pairs.

\begin{listing}
    \begin{sublisting}{\linewidth}
\begin{lstlisting}[style=sparqlStyle]
@prefix obo: <http://purl.obolibrary.org/obo/> .
@prefix rdfs: <http://www.w3.org/2000/01/rdf-schema#> .
obo:RO_0000056 rdfs:label "participates_in" .
[] a obo:CHEBI_53289 ;
    obo:RO_0000056 [ a pubchem:MeasureGroup ] ;
...
\end{lstlisting}
    \end{sublisting}
    \begin{sublisting}{\linewidth}
\begin{lstlisting}[style=tupleStyle]
('obo:CHEBI_53289', 'http://purl.obolibrary.org/obo/RO_0000056', 'participates_in', 'http://rdf.ncbi.nlm.nih.gov/pubchem/vocabulary#MeasureGroup')
...
\end{lstlisting}
    \end{sublisting}
    \hfill
    \begin{sublisting}{\linewidth}
\begin{lstlisting}[style=tupleStyle]
Instances of class 'obo:CHEBI_53289' have property 'obo:RO_0000056' (participates_in) with value type 'pubchem:MeasureGroup'.
...
\end{lstlisting}
    \end{sublisting}
    \caption{Formats to describe properties and value types used by instances of a class: Turtle (top), tuple (middle), natural language (bottom)}
    \label{list:embedding-cf}
\end{listing}

\subsection{Competency Question Generation}
\label{subsecsec:pipeline_descr-2}

This step invokes a language model to generate CQs based on various types of information about the KG: name and description, endpoint URL, list of the used ontologies. This information is either taken from the KG configuration (created in the previous step) or manually entered in a form.
The user may also provide any other relevant information, e.g. the abstract of an article describing the KG.

\begin{figure}[!t]
\includegraphics[width=\linewidth]{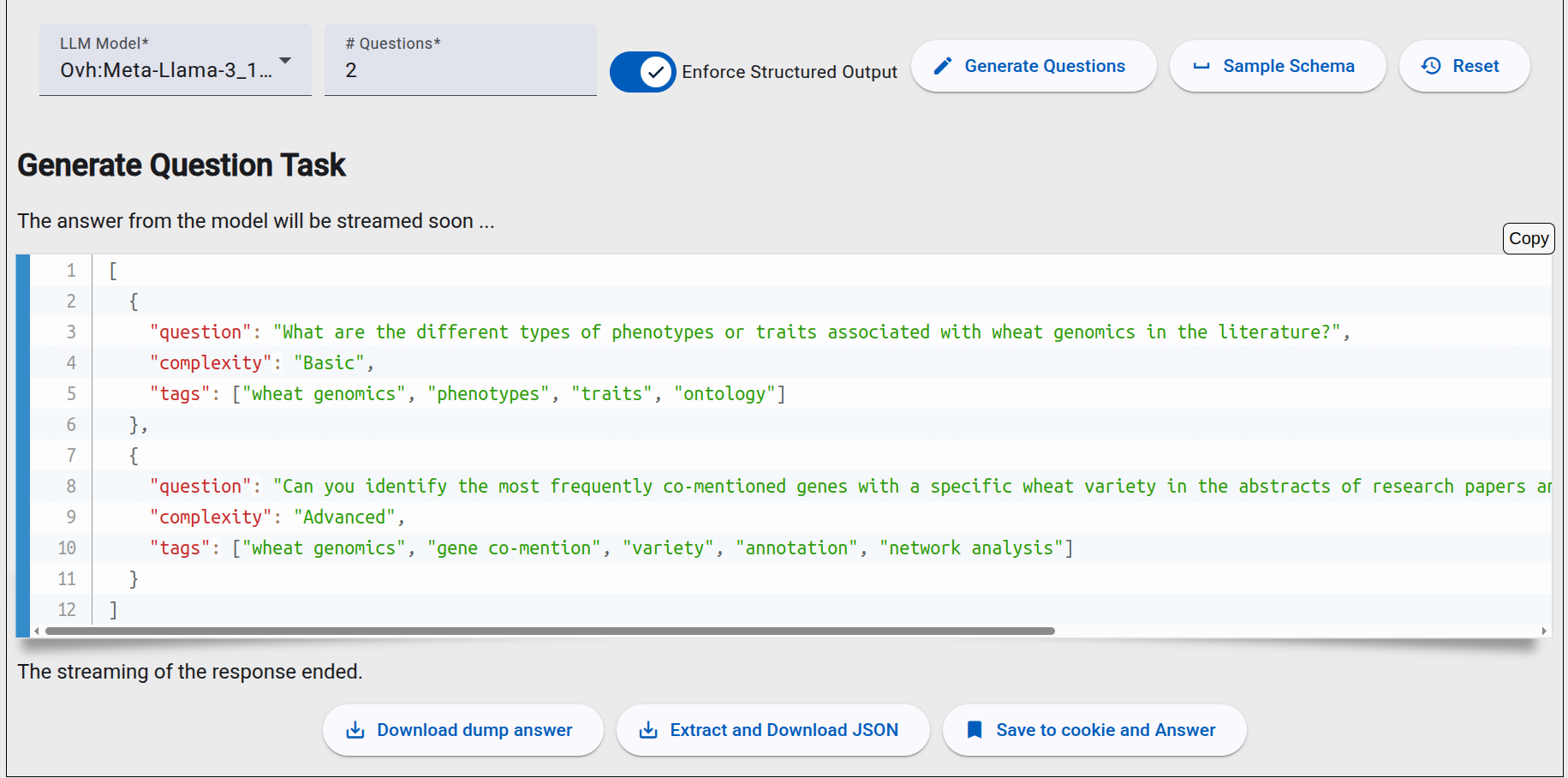}
\caption{The competency question generation interface.} \label{fig:df_cq_generation}
\end{figure}

Figure~\ref{fig:df_cq_generation} depicts the CQ generation interface.
The user can select the language model to be used, and the number of CQs to be generated. The model is instructed to return each question with an evaluation of its complexity (Basic, Intermediate or Advanced) and a set of tags.
The \textit{Enforce Structured Output} toggle can be used to compel the model to return the CQs as a JSON-formatted document.

Upon completion of the process, the user may download the output as a JSON document,
and save it in a browser's cookie for reuse in the next step.

\subsection{SPARQL Query Generation/Execution}
\label{subsecsec:pipeline_descr-3}

In this step, the user has the ability to generate a SPARQL query counterpart of a NL question, execute it against a KG, and get an interpretation of the results.
The question may originate from the preceding task, or the user may paste a question either hand-crafted or generated by another system.

\begin{figure}[!t]
\includegraphics[width=\linewidth]{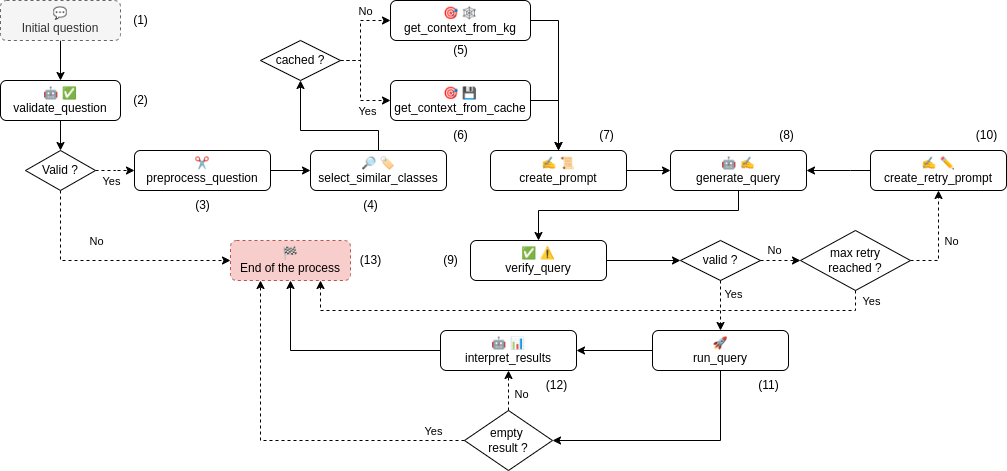}
\caption{\genkgbot SPARQL Query Generator/Executor: workflow of Scenario~5}
\label{fig:scenario_5_emojies}
\end{figure}

\begin{figure}[!t]
\centering
\includegraphics[width=\linewidth]{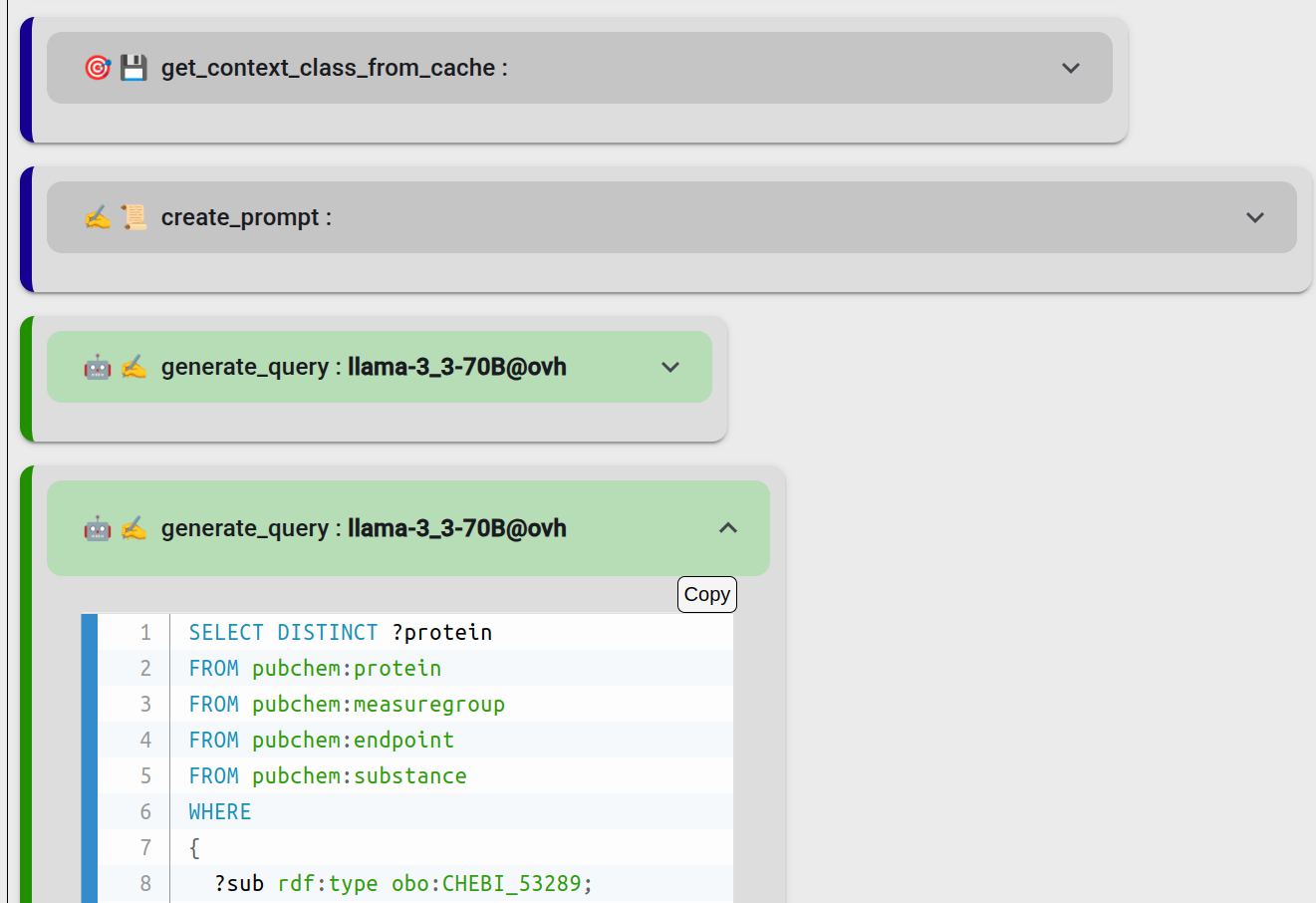}
\caption{The SPARQL Query Generator/Executor interface.} \label{fig:df_qa}
\end{figure}

\resourceName relies on various strategies provided by \genkgbot to accomplish this task, which we refer to as ``scenarios''.
In the following, we focus on Scenario~5, depicted in Figure~\ref{fig:scenario_5_emojies}, that we will further describe below.
When running a scenario, the steps of that scenario are progressively rendered on the interface, and for the ones that make an LLM call, the response is dynamically streamed to ensure a good user experience.
Figure~\ref{fig:df_qa} is a snapshot of the interface of the SPARQL Query Generator/Executor.
The steps are as follows:
\begin{enumerate}
    \item \textbf{Initial question:} the workflow is initiated by the user posing a NL question.
    \item \textbf{Question validation:} the question is evaluated by an LLM to assess the relevance of the question wrt. the description of the KG. The expected answer is boolean. If it is deemed invalid, the workflow stops.
    \item \textbf{Question pre-processing:} common techniques are used to extract named entities (NEs) from the question. SpaCy is used by default as it is widely adopted and considered production-class software. However, our implementation can support other NE extraction tools.
    \item \textbf{Select similar classes:} similarity search between the question and the ontology class descriptions computed in the KG pre-processing step is used to select relevant classes.
    \item \textbf{Get context information about the classes:} retrieve a description of the properties and value types used with instances of the selected classes.
    \item \textbf{Generate query:} generate a prompt from a template\footnote{\url{https://github.com/Wimmics/gen2kgbot/blob/master/app/scenarios/scenario_5/prompt.py\#L3}} using the KG configuration and the inputs from the previous steps, and submit it to the configured LLM.
    \item \textbf{Verify query and retry:} check if a SPARQL query was generated and if it is syntactically correct. If not, 
    generate a retry prompt that includes the last generated answer and the reason for the retry, e.g. syntax errors, and submit this retry prompt to the configured LLM.
    \item \textbf{Execute the SPARQL query:} if a valid SPARQL query was generated, submit it to the KG endpoint and get the results.
    \item Use the configured LLM to \textbf{interpret the SPARQL results}.
\end{enumerate}

Scenario~5, described above, is useful as a starting point when no prior question-query pair exists. However, once some pairs have been validated or if some pairs were hand-crafted, they can be added to the context and serve as examples. 
Scenario~6 can then be applied instead, as it provides the model with relevant example SPARQL queries that can help in generating more accurate queries with fewer refinement iterations.

\subsection{SPARQL Query Refinement}
\label{subsecsec:pipeline_descr-4}

\begin{figure}[!t]
\centering
\includegraphics[width=\linewidth]{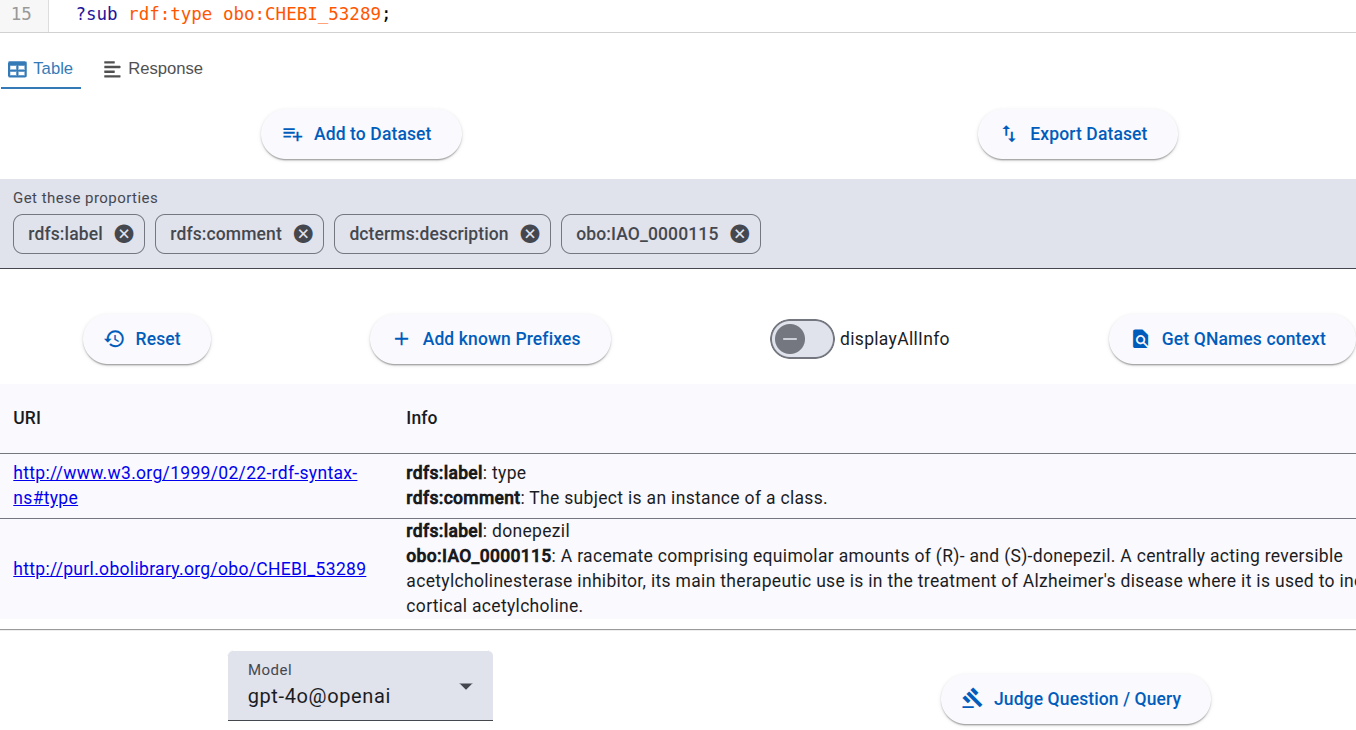}
\caption{The query refinement interface.} \label{fig:df_qr}
\end{figure}

In this step, the user can incrementally refine a SPARQL query so that it reflects precisely the question.
Figure~\ref{fig:df_qr} is a snapshot of the interface, and the process is as follows:

\begin{enumerate}
    \item First, the query is displayed in a SPARQL editor that highlights potential syntactic errors, and that can be used to submit the query to the endpoint.
    \item To help the user understand the query, \resourceName can extract the qualified (prefixed) names (QNs) and fully qualified names (FQNs) from the query and get their labels and descriptions.
    This is particularly useful with URIs that contain a non-significant identifier.
    For instance, the label of \url{http://purl.obolibrary.org/obo/CHEBI_53289} is ``donepezil'', and its description is ``a racemate comprising equimolar amounts of (R)- and (S)-donepezil (...)''.
    \item Then the LLM is asked to judge whether the query matches the given question. It is requested to provide a grade between 0 and 10 along with explanations justifying the grade.
\end{enumerate}
The user may then iterate as needed: amend the query based on the grade and insights from the model, test it, have the model judge it, etc.
Once a satisfying query is reached, the user can add the question-query pair to a dataset and export it in a variety of formats, catering to different use cases.

\section{Application Use Cases}
\label{sec:use_case}
In this section, we present several application use cases where \resourceName can be of great help. 
First, we focus on the creation of \qsets and the benefits of generating them instead of performing this task manually.
Second, we explore the application of \resourceName in creating a golden dataset for benchmarking, training and testing question answering models.
Finally, we examine how \resourceName can document existing KGs with multiple competency questions.

\textbf{Lowering the entry barrier to query rich KGs:}
Large public KGs usually provide lay users with user-friendly user interfaces that propose pre-defined queries and exploration options. Yet such interfaces can hardly accommodate complex custom queries where SPARQL expertise is necessary.
For instance, with over 111 million chemical substances and extensive bio-activity data, PubChem presents significant navigation challenges for researchers. Metabolomics experts often struggle to formulate complex SPARQL queries that would help them identify relationships between compounds, biological activities, and disease associations.
\resourceName addressed this challenge by allowing chemists to generate natural language questions and automatically convert them to SPARQL queries, thus drastically reducing the time spent on data retrieval.
To do so, researchers must provide a textual description of the KG together with additional relevant textual information, such as the abstracts of articles published about PubChem or about research made possible through PubChem.
For example, a researcher might ask: ``Which compounds have been tested against SARS-CoV-2 Main Protease and reported an IC50 below 1 $\mu$M?'' or ``Which natural product compounds from marine sponges show antimicrobial activities against \textit{Pseudomonas aeruginosa}?''.

Similarly, environmental health researchers studying the exposome face difficulties extracting meaningful correlations between environmental factors and metabolic responses across heterogeneous datasets. Currently, they rely either on simple predefined queries or must collaborate with knowledge engineering specialists, creating bottlenecks in research workflows. \resourceName could enable them to independently generate appropriate question-query pairs that bridge environmental exposures and biological outcomes, eliminating technical barriers to knowledge discovery. A typical researcher might need to ask: ``Which air pollutants are known to increase \textit{Nrf2} anti-oxydant protein expression?'' or ``What metabolic biomarkers show significant alterations following chronic exposure to per- and polyfluoroalkyl substances (PFAS) in human biomonitoring studies?''

\textbf{Ground truth and question-query benchmarks:} The Semantic Web community has pioneered challenges and benchmarks for question-answering over linked data~\cite{QALD10,DBNQA,LcQuAD}. However, each edition of these challenges requires updating \qsets for tasks that were proposed in previous editions, and creating new \qsets for newly proposed tasks. There does not exist a large number of such readily available \qsets, and they are often based on the same KGs (e.g. DBpedia, Wikidata). Setting up a new edition of a challenge 
therefore requires a significant effort to generate or update the training and test data.
\resourceName was designed to help produce these \qsets and can be used to facilitate the renewal of tasks for challenges and benchmarks on a variety of KGs.
For instance, the QALD Challenge~\cite{QALD10} has long been centered on DBpedia. In the latest edition, it was extended to Wikidata. Using \resourceName, we could extend the challenge with tasks targeting domain-specific graphs such as Uniprot~\cite{10.1093/nar/gky1049} or the aforementioned PubChem graph.

\textbf{Documenting a KG with competency questions:}
CQs are commonly used to demonstrate the basic capabilities of a KG. This requires working with domain experts to identify the CQs they may want to ask. In our experience, this is a time-consuming task involving multiple iterations.
\resourceName can be used to initialize, expand and enhance the scope and variety of CQs by systematically generating hundreds of competency questions.
For instance, PubChem's documentation currently provides a valuable foundation of 16~CQs.\footnote{\url{https://pubchem.ncbi.nlm.nih.gov/docs/rdf-use-cases}}
\resourceName could enhance this foundation by showcasing the full scope and complexity of chemical, biological, and pharmacological relationships within this extensive KG. 
The generated question sets can serve multiple purposes: providing entry points to new users, supporting KGs indexing, benchmarking search capabilities, identifying promising research directions, and accelerating the development of next-generation retrieval systems. 


\section{Preliminary Experimentations}
\label{sec:impact_reuse}

We have already identified three families of users, that correspond to the three uses cases described in Section~\ref{sec:use_case}: (1) the developers and maintainers of question answering systems, chatbots, conversational agents and other natural language search engines over KGs. The methods behind these systems all require to have \qsets to train, test and evaluate the system. They are the primary target of \resourceName.
(2) Events and groups organizing challenges, benchmarking existing solutions and building surveys. These are in constant need for new and renewed \qsets to compare the latest methods and establish the state-of-the-art. Here, \resourceName facilitates the creation of \qsets from any KG in any domain.
(3) While it is strongly recommended to document existing datasets and query services with examples of typical questions and queries, this is rarely done and, when it is, rarely extensive. \resourceName was designed to help KG publishers and maintainers to generate these examples with quality and quantity in mind.

We have initiated experiments with the first family of users on the IDSM KG related to the chemistry and metabolomics domain. 
Pharmaceutical researchers developing drug discovery platforms require comprehensive question-query pairs to train intelligent systems for identifying promising molecular candidates across multiple parameters. These researchers benefit from \resourceName's ability to generate diverse questions exploring structure-activity relationships, pharmacokinetic properties, and target binding profiles. 
Metabolomics data scientists integrating multi-omics datasets need sophisticated query templates that traverse complex biochemical pathway knowledge, particularly when correlating mass spectrometry findings with biological outcomes. 
Based on this experimentation, we believe that academic laboratories focusing on cheminformatics and bioinformatics can utilize \resourceName to develop educational materials demonstrating how semantic queries extract meaningful insights from chemical databases. 
\resourceName can significantly reduce technical barriers that have historically prevented domain experts from fully leveraging KG technologies in their specialized fields.

For the third family of users, we have experimented \resourceName with outputs of the \href{http://www.d2kab.org/}{D2KAB} project. D2KAB produced several datasets among which the \textit{Wheat Genomics Scientific Literature Knowledge Graph}~\cite{yacoubiayadi:hal-04495022} that represents the named entities extracted from a corpus of over 8,000 PubMed articles related to wheat genetics and genomics.
The NEs include genes, phenotypes, taxon names and varieties in titles and abstracts.
During the project, we worked with domain experts to figure out several CQs\footnote{\url{https://github.com/Wimmics/WheatGenomicsSLKG/blob/main/SPARQLQueries-JupyterNotebook.ipynb}} to document the graph and illustrate its usefulness.
When tested with this KG, \resourceName was able to automatically generate relevant CQs and translate them into SPARQL queries that were close to the target. After a short refinement step, we managed to get valid question-query pairs. Based on our initial experiments with these two domain-specific KGs, Q²Forge shows promise for reuse across different contexts, though comprehensive evaluation remains as important future work.

Our experiments on the IDSM and D2KAB KGs allowed us to determine the time required to execute each stage of the pipeline. Table ~\ref{tab:experiments} summarises statistics for each KG, as well as the time taken to (1) compute classes' text embeddings, (2) generate 50 CQs, (3) answer one CQ, (4) extract QNs and FQs and obtain one refinement proposal. The experiments were performed using: an Intel Core Ultra 9 185H × 22 CPU with 64GB of RAM and an NVIDIA RTX 2000 Ada Generation Laptop GPU (8GB). We used \texttt{nomic-embed-text} embedding model and the FAISS vector store for the embedding task, and DeepSeek-v3 seq2seq model for all LLM calls.


\begin{table}[h!]
\resizebox{\linewidth}{!}{  
\begin{tabular}{cccccccc}
\textbf{KG} & \textbf{\#triple} & \textbf{\#cls} & \textbf{\#ppt} & \textbf{(1)} & \textbf{(2)} & \textbf{(3)} & \textbf{(4}) \\ \hline
D2KAB & 27,093,602 & 590 & 287 & 10 & 132.4 & 45.6 & 29.2 \\ \hline
IDSM  & 36,285,192,866 & 226,809 & 1,044 & 2,592.5 & 146.2 & 51 & 40.1 \\ \hline
\end{tabular}
}
\vspace{0.25cm} 
\caption{IDSM and D2KAB: statistics and time requirements in seconds for \resourceName steps to execute.}
\label{tab:experiments}
\end{table}


\section{Source Code and Documentation}
\label{sec:code_doc}

\resourceName and accompanying \genkgbot integrate several software components that are robust and have been proven effective in various contexts. LangChain and LangGraph\footnote{\url{https://www.langchain.com/}} are used for LLM workflow orchestration, Spacy\footnote{\url{https://spacy.io/}} for question pre-processing and named entity recognition, rdflib\footnote{\url{https://github.com/RDFLib}} for the manipulation of RDF data, and YasGUI\footnote{\url{https://yasgui.triply.cc/}} as a SPARQL editor.

\textbf{Source Code Availability.} 
\resourceName and \genkgbot are provided under the GNU Affero General Public License v3.0 or later (AGPL-3.0-or-later) license. The code is published on public Github repositories, 
and the versions used at the time of writing are identified by DOIs to ensure the long-term preservation and citability.\footnote{All links are given at the beginning of this article.}
A prototype is available for public access and has been assigned a W3ID. The API provided by \genkgbot is documented according to the OpenAPI format.\footnote{\url{http://w3id.org/q2forge/api/docs/}}
Table~\ref{tab:recap-links} summarises the links to the source code, demonstration videos and online prototype of \resourceName.

\textbf{Sustainability Plan.} Over the next four years, financial support has been secured through the MetaboLinkAI project.\footnote{\url{http://www.metabolinkai.net/}} This project aims to transform metabolomics data into actionable insights through the utilisation of AI-powered, knowledge graph-driven solutions. 
This will provide an opportunity to evaluate the quality, relevance and applicability of \resourceName in the chemistry domain.
Moreover, a fundamental objective of \resourceName is to provide a generic solution, reusable with a variety of KGs. Consequently, we intend to provide support to communities expressing interest and willingness to experiment with it for their own needs. 
This support may range from best-effort to more formalized collaboration. To support collaborations, adoptions and contributions we secured two other contributors: the P16 public program\footnote{\url{https://p16.inria.fr/en/}}  that helps open-source project improve their code and diffusion, and the Probabl company\footnote{\url{https://probabl.ai/about}} whose mission is to develop, maintain the state-of-the-art, sustain, and disseminate a complete suite of open source tools for data science. 

\begin{table}[h!]
\centering
\resizebox{\linewidth}{!}{  
\begin{tabular}{lll}
\textbf{License}                              & GNU Affero General Public License v3.0 or later (AGPL-3.0-or-later) \\ \hline
\textbf{Online prototype}                     & \url{https://www.w3id.org/q2forge/} \\ \hline
\textbf{\resourceName}   & \textbf{Repo} \url{https://github.com/Wimmics/q2forge}               
                       \textbf{DOI} \href{https://doi.org/10.5281/zenodo.15388693}{10.5281/zenodo.15388693} \\  
\textbf{\genkgbot} & \textbf{Repo} \url{https://github.com/Wimmics/gen2kgbot} \textbf{DOI} \href{https://doi.org/10.5281/zenodo.15388687}{10.5281/zenodo.15388687}         \\ \hline
\textbf{Demo video} & \textbf{Teaser} \url{https://youtu.be/E9rgCZzWH4k} \textbf{Full} \url{https://youtu.be/I3w-jmZRJII} \\ \hline
\end{tabular}%
}
\vspace{0.25cm} 
\caption{License and links to the source code, demonstration videos and online prototype of \resourceName.}
\label{tab:recap-links}
\end{table}


\section{Conclusion and Perspectives}
\label{sec:conclusion}
The present article has highlighted the challenge of creating \qsets (datasets of question-query pairs) for a KG as part of the ever-growing LOD cloud. 
To address this challenge, concrete methods and tools have been presented. 
Utilising robust, industry-proven tools, \resourceName's pipeline is designed to address the generation of competency questions (CQ) in NL, translate the CQs into SPARQL queries, and help users to refine those queries, and export high-quality QALD-like \qsets that can be used for benchmarking, training and evaluating text-to-SPARQL models.

This innovative end-to-end pipeline incorporates a variety of dedicated services to achieve these objectives. It emphasizes genericity (it can apply to any KG in any domain), extensibility (the pipeline can easily be modified and extended to account for new needs), and flexibility
(the various tasks of the pipeline can be executed as a whole, and some tasks can be used independently of the others).
In addition to the interfacing with third-party systems shown in Figure~\ref{fig:pipeline}, the system is designed to be reused and integrated into other scenarios. In education for instance, when teaching SPARQL, \resourceName could be modified to serve as a tailored instructor guiding learners to navigate the complexities of SPARQL. Furthermore, since the query refinement task can be accessed independently of the other tasks (its URL takes arguments ``question'' and ``query''), adding the appropriate button to an existing SPARQL editor could seamlessly integrate this task into an existing workflow.
Furthermore, although the pipeline does not explicitly address multilingualism, multilingual CQ generation is feasible with minimal modifications (depending on the LLM's capabilities) since the configuration-driven architecture supports language-specific prompt templates. Yet, for optimal results, ontology descriptions and metadata may need translation into the target language.

Current development of protocols such as MCP (Model Context Protocol),\footnote{\url{https://modelcontextprotocol.io/}} A2A (Agent-to-Agent)\footnote{\url{https://google.github.io/A2A/}} and hMAS (Hypermedia Multi-Agent Systems)\footnote{\url{https://project.hyperagents.org/}} reflects ongoing efforts to simplify integration, enhance collaboration, and ensure secure and efficient communication between AI agents and external systems. 
These protocols can potentially be interfaced with \resourceName to facilitate its incorporation into broader systems. 
In particular, we plan to implement MCP to expose the three primary functions of \resourceName as reusable tools, enabling seamless integration of \resourceName's components into other workflows. 
Conversely, \resourceName could be extended to support the invocation of MCP servers providing access to third-party services such as knowledge graphs.
Additionally, incorporating A2A would allow \resourceName to support multi-agent collaboration across diverse ecosystems, fostering coordination between agents of varying frameworks. Finally, aligning with hMAS would leverage semantic hypermedia for uniform interactions among people, devices, and digital services, creating hybrid AI communities that operate transparently and accountably on the Web. These extensions would make \resourceName even more versatile, facilitating the development of KG applications in different domains.

Besides the implementation of MCP mentioned above, future work will proceed along several directions in the short and medium term.

First, we will focus on data quality evaluation. This includes establishing automated validation processes and human evaluation protocols to ensure the relevance and accuracy of generated question-query pairs. In addition, we will evaluate \resourceName’s sensitivity to different LLM choices and analyze the resulting trade-offs in cost and latency. We will also identify and characterize the failure types in \genkgbot across a spectrum of challenging queries using the variety of operators available in SPARQL.

Second, we envision integrating existing KG construction and extension tools from unstructured content.
Such integration would enable \resourceName to generate questions about both structured and unstructured data sources, broadening its applicability across diverse data landscapes.
%
%
Third, we wish to address several current limitations: 
(1) Follow-up questions: our plan is to achieve this using short/long memory and conversation summarizing to cope with context explosion; 
(2) custom and tailor result visualization when a textual rendering of a SPARQL query result is not the optimum way of conveying the results; 
(3) Federated queries: the current implementation is limited to single SPARQL endpoints and does not support federated queries across multiple KGs, which restricts its applicability in scenarios requiring data integration from diverse sources. This limitation could be addressed by extending \resourceName architecture to support the generation of queries that span multiple endpoints
using the SERVICE clause of the SPARQL 1.1 Federation specification. 
Finally, extensive dissemination and open-source support activities are planned so that other communities can adopt this work and adapt it to their specific needs. Through community engagement, documentation, and targeted outreach, we aim to foster broader adoption and enable customization of \resourceName across various domains and research contexts.

\begin{acks}
This work was supported by the French government through the France 2030 investment plan managed by the National Research Agency (ANR), as part of the Initiative of Excellence Université Côte d'Azur (ANR-15-IDEX-01). Additional support came from French Government's France 2030 investment plan (ANR-22-CPJ2-0048-01), through 3IA Cote d'Azur (ANR-23-IACL-0001) as well as the MetaboLinkAI bilateral project (ANR-24-CE93-0012-01 and SNSF 10002786).
\end{acks}

\bibliographystyle{ACM-Reference-Format}
\bibliography{biblio}

\end{document}